\documentclass[reprint,superscriptaddress,
amsmath,amssymb,
aps,
prl,
]{revtex4-2}

\usepackage[pdftex]{graphicx}
\usepackage{dcolumn}
\usepackage{bm}
\usepackage[pdftex]{hyperref}

\begin{document}
  \title{Skyrmion and vortex crystals in the Hubbard model}
  \author{Kaito Kobayashi}
  \affiliation{Department of Applied Physics, the University of Tokyo, Tokyo 113-8656, Japan}
  \author{Satoru Hayami}%
  \affiliation{Graduate School of Science, Hokkaido University, Sapporo 060-0810, Japan}

  \date{\today}

  \begin{abstract}
    A mutual interplay between the charge and spin degrees of freedom in itinerant magnets leads to a plethora of topological spin textures, such as magnetic skyrmion and vortex crystals, in both centrosymmetric and noncentrosymmetric hosts. Meanwhile, their stabilization has been extensively studied in the system including the classical localized spins. We here study a realization of the skyrmion crystal in the centrosymmetric triangular-lattice Hubbard model, where the itinerant nature of electrons plays a more significant role. By performing the self-consistent mean-field calculations, we find that two types of skyrmion crystals with spatially nonuniform charge modulations appear in the ground state at a zero magnetic field. Moreover, we obtain another noncoplanar vortex crystal phase without a net scalar chirality in the vicinity of the skyrmion crystal phase. We show that the latter vortex crystal exhibits the topological phase transition to a different skyrmion phase in an applied magnetic field. Our results provide a possibility of the skyrmion and vortex crystals in itinerant magnets without the localized moments.
\end{abstract}

\maketitle

Itinerant magnets manifest themselves not only in various magnetic orderings but also in unusual conductive phenomena as a consequence of the synergy between charge and spin degrees of freedom in electrons. 
The microscopic mechanisms of magnetically-ordered phases are often different from those in insulating magnets; an effective interaction through the kinetic motion of electrons as well as the Coulomb interaction plays an important role in determining optimal spin configurations. 
A typical example is the Ruderman-Kittel-Kasuya-Yosida (RKKY) interaction~\cite{1954:PR:RK,1956:PTP:K,1957:PR:Y}, which leads to the instability toward helical magnetic ordering with a long period structure featured by a single-\(Q\) modulation. Furthermore, such a spin-charge entanglement brings about a comparable higher-order multiple-spin interaction to the RKKY interaction depending on the electronic band structures, which results in noncoplanar magnetic orderings characterized by a superposition of the single-$Q$ helical states, i.e., a multiple-$Q$ state~\cite{2008:PRL:MatinBatista,2010:Akagi:JPSJ,Chern:PhysRevLett.105.226403,2012:PRL:Akagi,Solenov:PhysRevLett.108.096403,Hayami:PhysRevB.90.060402,Barros:PhysRevB.90.245119,Ozawa:doi:10.7566/JPSJ.85.103703,batista2016frustration,wang2021skyrmion}. 
These nontrivial spin textures are the source of nontrivial conductive phenomena even without relativistic spin-orbit coupling, such as the topological Hall effect~\cite{2000:PRB:Ohgushi, Shindou:PhysRevLett.87.116801,2007:JPCM:Sinitsyn,2009:PRL:Neubauer,2009:PRL:Lee, Nagaosa:RevModPhys.82.1539, Xiao:RevModPhys.82.1959} and the nonreciprocal transport~\cite{2018:Ncomm:Tokura,2020:PRB:Hayami,hayami2022nonlinear,hayami2022nonreciprocal}.
  
Recently, it was revealed that a magnetic skyrmion crystal (SkX), which has a topologically protected noncoplanar spin texture, is also realized by the itinerant nature of electrons based on the analysis for the Kondo lattice model consisting of itinerant electrons and classical localized spins~\cite{2017:PRL:Ozawa,hayami2021topological}. 
Subsequently, the emergence of the SkX is accounted for by an effective spin model with the positive biquadratic interaction in addition to the RKKY interaction, which originates from the Fermi surface instability~\cite{2017:PRB:Hayami}. 
As this mechanism can be applied to any lattice structures irrespective of spatial inversion symmetry, it might provide the microscopic origin of the SkXs observed in centrosymmetric magnets, such as \(\mathrm{Gd_2PdSi_3}\)~\cite{2019:Science:Kurumaji,sampathkumaran2019report,Nomoto:PhysRevLett.125.117204,paddison2022magnetic,Bouaziz:PhysRevLett.128.157206}. 
Indeed, magnetic phase diagrams in centrosymmetric skyrmion-hosting materials have been reproduced by the effective spin model, such as GdRu$_2$Si$_2$~\cite{khanh2020nanometric,Yasui2020imaging,khanh2022zoology}, Gd$_3$Ru$_4$Al$_{12}$~\cite{hirschberger2019skyrmion,Hirschberger:10.1088/1367-2630/abdef9}, and EuAl$_4$~\cite{Shang:PhysRevB.103.L020405,takagi2022square,hayami2022multiple,Zhu:PhysRevB.105.014423}. 

Meanwhile, there is a natural question about whether the SkX remains robust or not when electrons show the itinerant nature rather than the localized one.
In other words, it is unclear whether the instability toward the SkX occurs by the kinetic motion of electrons even without the classical spin degree of freedom. 
To answer this question, we investigate the Hubbard model without the localized spins. 
Based on the self-consistent mean-field calculations for the Hubbard model on a centrosymmetric triangular lattice, we show that two types of the SkX appear even without the external magnetic field, as found in the Kondo lattice model~\cite{2017:PRL:Ozawa}, although the spin amplitudes are much smaller and strongly depend on the site. 
Besides, we discover a triple-$Q$ (3$Q$) vortex phase without the spin scalar chirality that has never been reported as the ground state in the Kondo lattice model. 
We show that the magnetic field causes the topological phase transition from the 3$Q$ vortex phase to another SkX phase. 
Our results indicate the importance of the charge degree of freedom, which becomes a source of not only the SkX but also different types of the vortex phase in itinerant magnets.   

Let us consider the Hubbard model on a two-dimensional triangular lattice, which is given by 
\begin{equation}
\label{eq: Ham}
  \mathcal{H}=-\sum_{ij\sigma}t_{ij}c_{i\sigma}^\dagger c_{j\sigma}+U\sum_{i}n_{i\uparrow}n_{i\downarrow}-B\sum_{i}S^z_i
\end{equation}
where the operator \(c_{i\sigma}^\dagger\) (\(c_{i\sigma}\)) is a creation (annihilation) operator for an electron with spin $\sigma$ at site \(i\).  
The first term describes the kinetic energy of the electrons with the transfer integral between sites \(i\) and \(j\), \(t_{ij}\). The second term represents the onsite Coulomb interaction with the coupling constant $U$ where $n_{i\sigma}=c^\dagger_{i\sigma}c^{}_{i\sigma}$. The third term represents the Zeeman coupling to an external field along the $z$ direction; \(S_i^z= \frac{1}{2}(c^{\dagger}_{i\uparrow}c^{}_{i\uparrow}-c^{\dagger}_{i\downarrow}c^{}_{i\downarrow})\).

In the following calculations, we consider the nearest-neighbor hopping, \(t_1=1\), and third-neighbor hopping, \(t_3=-0.85\), since these hopping parameters lead to the Fermi-surface instability at the commensurate wave vectors, \(\bm{Q}_1=(\pi/3,0)\), \(\bm{Q}_2=(-\pi/6,\sqrt{3}\pi/6)\), and \(\bm{Q}_3=(-\pi/6,-\sqrt{3}\pi/6)\) for the electron filling $n_{i}\equiv \sum_{\sigma}n_{i\sigma}\simeq
0.398$ (the lattice constant of the triangular lattice is set as unity); the instability toward the SkX with the 3$Q$ modulation at $\bm{Q}_1$-$\bm{Q}_3$ has been found in the Kondo lattice model with the classical localized spins~\cite{2017:PRL:Ozawa}.

For the model in Eq.~(\ref{eq: Ham}), we perform the self-consistent mean-field calculations at zero temperature based on the Hartree-Fock approximation for the Coulomb interaction; 
\(\sum_{i}n_{i\uparrow}n_{i\downarrow} \simeq \langle n_{i\downarrow}\rangle n_{i\uparrow}+\langle n_{i\uparrow} \rangle n_{i\downarrow}-\langle n_{i\uparrow} \rangle\langle n_{i\downarrow}\rangle -\langle c^{\dagger}_{i\downarrow}c^{}_{i\uparrow} \rangle c^{\dagger}_{i\uparrow}c^{}_{i\downarrow}-\langle c^{\dagger}_{i\uparrow}c^{}_{i\downarrow} \rangle c^{\dagger}_{i\downarrow}c^{}_{i\uparrow}+\langle c^{\dagger}_{i\uparrow}c^{}_{i\downarrow} \rangle\langle c^{\dagger}_{i\downarrow}c^{}_{i\uparrow} \rangle\).

We consider the magnetically ordered states in the 48-site magnetic unit cell ($N_{\rm m}=48$) under the periodic boundary conditions to accommodate the SkX with $\bm{Q}_1$-$\bm{Q}_3$ [see also Fig.~\ref{fig2}(a) for instance].
We introduce the $N_{\bm{k}}=60\times 60$ supercells to reduce the finite-size effect, with almost the same results for $N_{\bm{k}}=32\times 32$.
The convergent condition is determined by 
\(\Delta(\langle c^\dagger_{i\sigma}c_{i\sigma'}\rangle)\)\(<10^{-8}\) for each site $i$ and spin $\sigma$.

Once the convergent spin configurations are obtained, we examine the spin and charge properties by calculating the spin and charge structure factors, $S(\bm{q})$ and $N(\bm{q})$, respectively, which are given by 
  \begin{align}
    S(\bm{q}) &= \frac{1}{N}\sum_{ij}\langle\bm{S}_i\rangle \cdot \langle \bm{S}_j \rangle e^{i\bm{q}\cdot (\bm{r}_i-\bm{r}_j)},\\
    N(\bm{q}) &= \frac{1}{N}\sum_{ij}\Delta n_i\Delta n_j e^{i\bm{q}\cdot (\bm{r}_i-\bm{r}_j)} ,
  \end{align}
where $\bm{r}_i$ is the position vector at site $i$ in the larger magnetic unit cell for \(N=12\times 12 = 3N_{\rm m}\) sites corresponding to a modulation period of \(\pi/6\); $\Delta n_i\equiv \sum_{\alpha}\langle c_{i\alpha}^\dagger c_{i\alpha}\rangle  - n_{\mathrm{ave}}$ with the average charge density \(n_{\mathrm{ave}}\). 
We also calculate the total magnetization per site \(M_z = \sum_{i} \langle S_i^z\rangle/N_{\rm m}\). 
In addition, to identify whether the obtained states are topologically trivial or not, we compute the scalar spin chirality per triangle in the magnetic unit cell, $\chi_{\mathrm{tot}}=\sum_{\{ijk\}}\langle\bm{S}_i\rangle\cdot(\langle\bm{S}_j\rangle\times \langle\bm{S}_k\rangle)/2N_{\rm m}$. 
The nonzero value of $\chi_{\mathrm{tot}}$ becomes the origin of the topological Hall effect~\cite{2000:PRB:Ohgushi}. 

\begin{figure}[t]
  \includegraphics[width=\hsize]{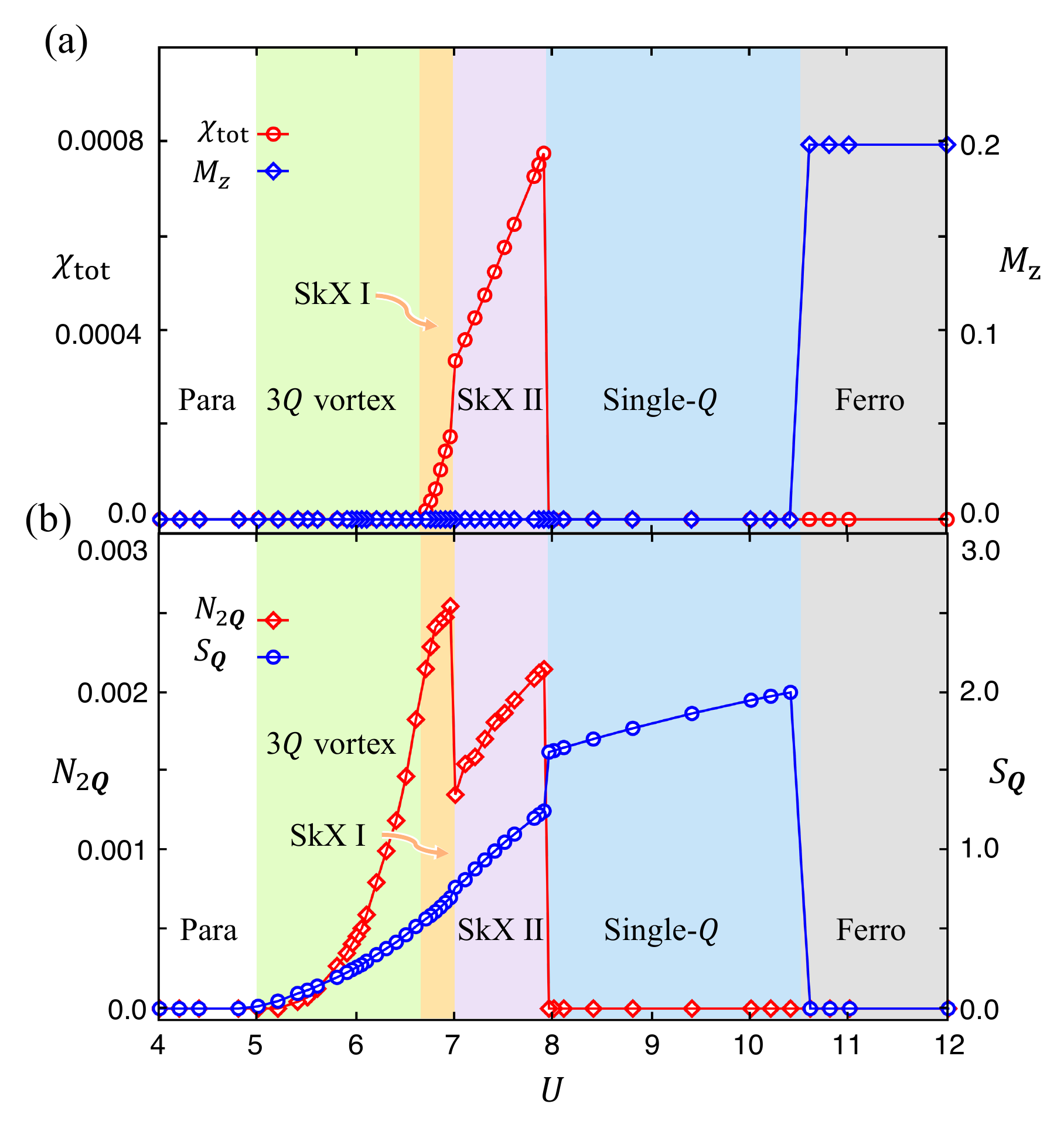}
  \caption{
  \(U\) dependence of (a) the total scalar chirality \(\chi_{\mathrm{tot}}\) and the magnetization \(M_z\), (b) the charge structure factor \(N_{2\bm{Q}}\) and the spin structure factor \(S_{\bm{Q}}\) summed over \(2\bm{Q}_{\mu}\) and \(\bm{Q}_{\mu}\), respectively. The different colors in the background represent the different phases.}
  \label{fig1}
\end{figure}
 
Figure~\ref{fig1} shows the ground-state phase diagram against $U$ at a zero field, i.e., $B=0$. There are six phases characterized by different spin, charge, and chirality quantities. We show the $U$ dependence of $\chi_{\mathrm{tot}}$ and $M_z$ in Fig.~\ref{fig1}(a) and the sum of the structure factors at the commensurate wave vectors $\bm{Q}_\mu$ and $2\bm{Q}_\mu$, \(N_{2\bm{Q}}\equiv \sum_{\mu}N(2\bm{Q}_{\mu})$ and $S_{\bm{Q}}\equiv \sum_{\mu}S(\bm{Q}_{\mu})\) in Fig.~\ref{fig1}(b). In the weak-(strong-)correlation regime, the paramagnetic (ferromagnetic) state is realized, whose tendency has been found in the Hubbard model~\cite{Hirsch:PhysRevB.31.4403,Claveau:2014}. Meanwhile, in the intermediate-correlation regime, we obtain four magnetic phases. Among them, three out of the four are characterized by the multiple-$Q$ states: the \(3Q\) vortex phase, SkX I, and SkX II. The remaining single-$Q$ state next to the ferromagnetic state is characterized by the single-$Q$ spiral state with \(S_{\bm{Q}}\) but without $\chi_{\mathrm{tot}}$, $M_z$, and $N_{2\bm{Q}}$. 

\begin{figure}[htbp]
  \includegraphics[width=\hsize]{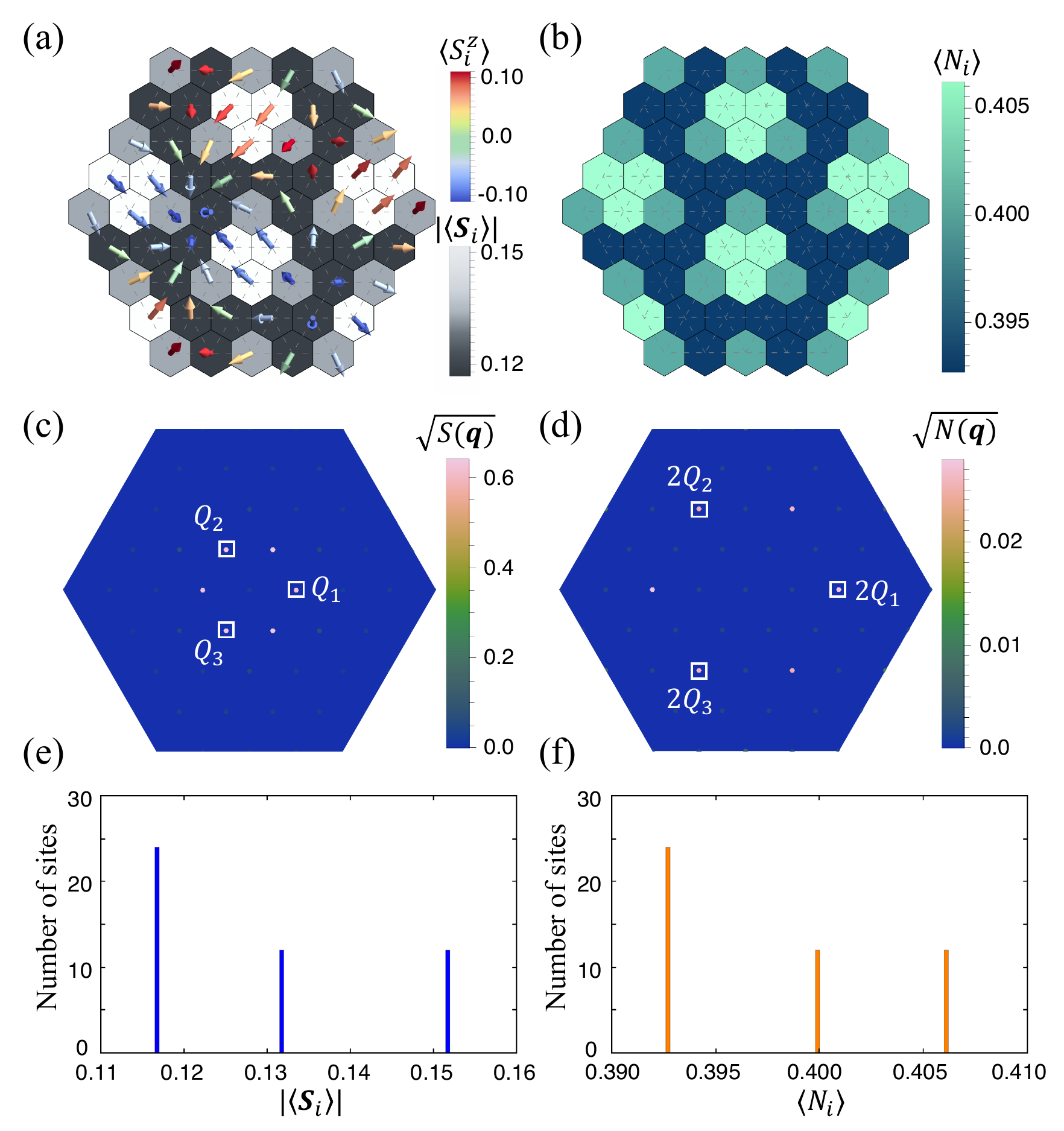}
  \caption{(a) Real-space spin configuration in the SkX II phase at \(U=7.8\). The arrows represent the spins, their colors display the \(z\)-spin component, and the contour shows the spin norm $|\langle \bm{S}_i\rangle |$. 
  (b) The contour plot for the charge density. (c),(d) The intensity plot of the square root of (c) the spin structure factor \(S(\bm{q})\) and (d) the charge structure factor \(N(\bm{q})\). (e) The histogram of $|\langle \bm{S}_i\rangle |$ with a bin width of \(0.0005\). (f) The histogram of  \(\langle N_i\rangle \) with a bin width of \(0.0002\).}
  \label{fig2}
\end{figure}

For $7.0 \lesssim  U \lesssim 7.9$, we find that the SkX II with nonzero $\chi_{\rm tot}$ but zero $M_z$ becomes the ground state, as shown in Fig.~\ref{fig1}(a). This state has triple-$Q$ charge and spin modulations, as shown in Fig.~\ref{fig1}(b). Indeed, almost the same intensities at $\bm{Q}_\mu$ ($2\bm{Q}_\mu$) are observed in the spin (charge) structure factor, as shown in Figs.~\ref{fig2}(c) and \ref{fig2}(d). It is noted that the triple-$Q$ peak structure in $N({\bm{q}})$ is driven by the triple-$Q$ spin density waves. The peak position at $2\bm{Q}_\mu$ indicates that the constituent spin density waves propagate in an orthogonal way in spin space~\cite{2021:PRB:Hayami, Eto:PhysRevLett.129.017201}. The equal triple-$Q$ intensities in $S(\bm{q})$ and \(N (\bm{q})\) appear in the real-space spin and charge distributions in Figs.~\ref{fig2}(a) and \ref{fig2}(b), respectively; both $|\langle \bm{S}_i\rangle |$ and $\langle N_i \rangle $ are distributed in a threefold-symmetric way. In addition, we find that both of them exhibit three discrete values, as shown in Figs.~\ref{fig2}(e) and \ref{fig2}(f), which means there are three independent environments in the SkX II phase. This phase presumably corresponds to the SkX stabilized in the Kondo lattice model~\cite{2017:PRL:Ozawa}. 

\begin{figure}[tbp]
  \includegraphics[width=\hsize]{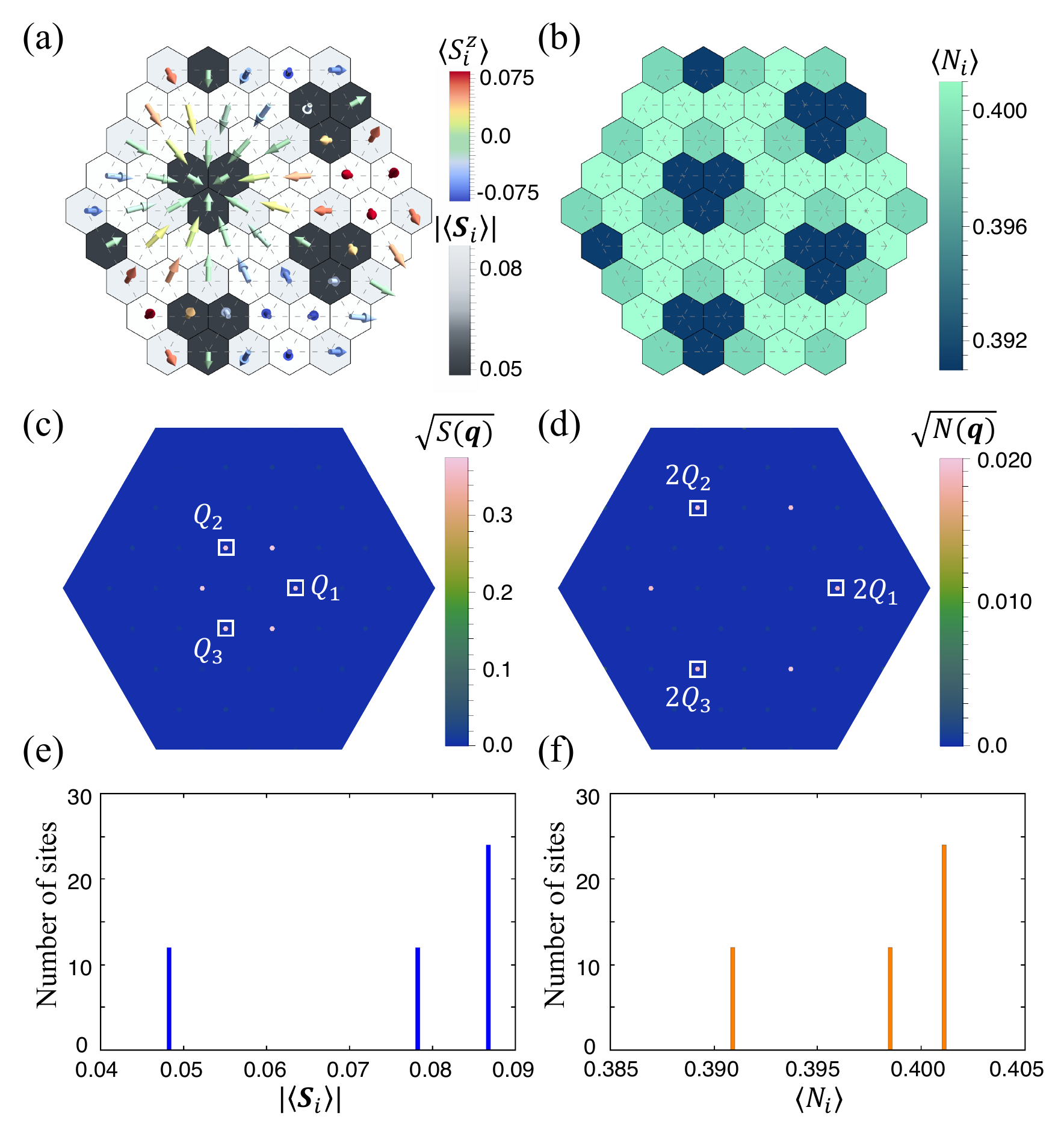}
  \caption{The same plots as in Fig.~\ref{fig2} in the 3$Q$ vortex phase at $U=6.4$}
  \label{fig3}
\end{figure}

On the other hand, for $5.0 \lesssim  U \lesssim 6.6$, the 3$Q$ vortex phase is stabilized, which also exhibits triple-$Q$ charge and spin modulations [Fig.~\ref{fig1}(b), Fig.~\ref{fig3}(c), and Fig.~\ref{fig3}(d)] but no $\chi_{\rm tot}$ and $M_z$ [Fig.~\ref{fig1}(a)]; the charge modulation at $2\bm{Q}_\mu$ indicates a similar multiple-$Q$ superposition to the SkX II. In addition, the absence of $\chi_{\rm tot}$ suggests the occurrence of the phase shift among the constituent waves from the SkX II~\cite{hayami2021phase,Hayami:PhysRevResearch.3.043158,Shimizu:PhysRevB.105.224405}, which has been found at finite temperatures in the Kondo lattice model~\cite{hayami2021phase}. The real-space spin configuration is characterized by four different spin vortices where the spin norm $|\langle \bm{S}_i\rangle |$ becomes small, as shown in Fig.~\ref{fig3}(a). Accordingly, the charge distribution is also threefold symmetric around the vortices with small $|\langle N_i\rangle |$, as shown in Fig.~\ref{fig3}(b). The histograms of $|\langle \bm{S}_i\rangle |$ [Fig.~\ref{fig3}(e)] and $\langle N_i \rangle $ [Fig.~\ref{fig3}(f)] show the opposite tendency to the SkX II in Figs.~\ref{fig2}(e) and \ref{fig2}(f); the sites with larger $|\langle \bm{S}_i\rangle |$ and $\langle N_i \rangle$ are preferred in the 3$Q$ vortex. The ratio of maximum and minimum values of $|\langle \bm{S}_i\rangle |$ is around \(0.56\), which is much smaller than that of the SkX II phase \(0.77\), while the ratio in terms of $\langle N_i \rangle$ is almost constant: \(0.975\) for the 3\(Q\) vortex and \(0.967\) for the SkX II. This result implies that the site-dependent moment reduction due to the kinetic motion of electrons, i.e., the charge degree of freedom, is important for the realization of the 3$Q$ vortex as the ground state.

In the narrow region sandwiched by the 3$Q$ vortex and the SkX II, the SkX I appears. Although this state shows a nonzero scalar chirality like the SkX II, its spin and charge structure factors appear to be continuously connected from the 3$Q$ vortex phase, as shown in Figs.~\ref{fig1}(a) and \ref{fig1}(b). Thus, the SkX I is regarded as the intermediate state between the 3$Q$ vortex and SkX II. 

Let us compare the present results with those in the Kondo lattice model~\cite{2017:PRL:Ozawa}. In both models, the SkXs characterized by the triple-$Q$ structures in $S(\bm{q})$ and $N(\bm{q})$ are stabilized at a zero field, but a distinct difference between them is found in the spin norm. 
The SkX II in the Hubbard model shows the strong spatial dependence of $|\langle \bm{S}_i \rangle|$ [Fig.~\ref{fig2}(a)] in contrast to the case of the Kondo lattice model with the fixed-length classical spins. 
This indicates that the itinerant character appears more clearly in the Hubbard model. Such a qualitative difference might 
be the origin of another triple-$Q$ vortex phase in Fig.~\ref{fig3}(a), which is not stabilized as the ground state in the Kondo lattice model due to the constraint in terms of the spin norm. 
Thus, the interplay between the charge and spin degrees of freedom plays an important role in inducing the noncoplanar multiple-$Q$ states, which will bring about further exotic spin-charge entangled topological spin textures. 

\begin{figure}[t]
  \includegraphics[width=\hsize]{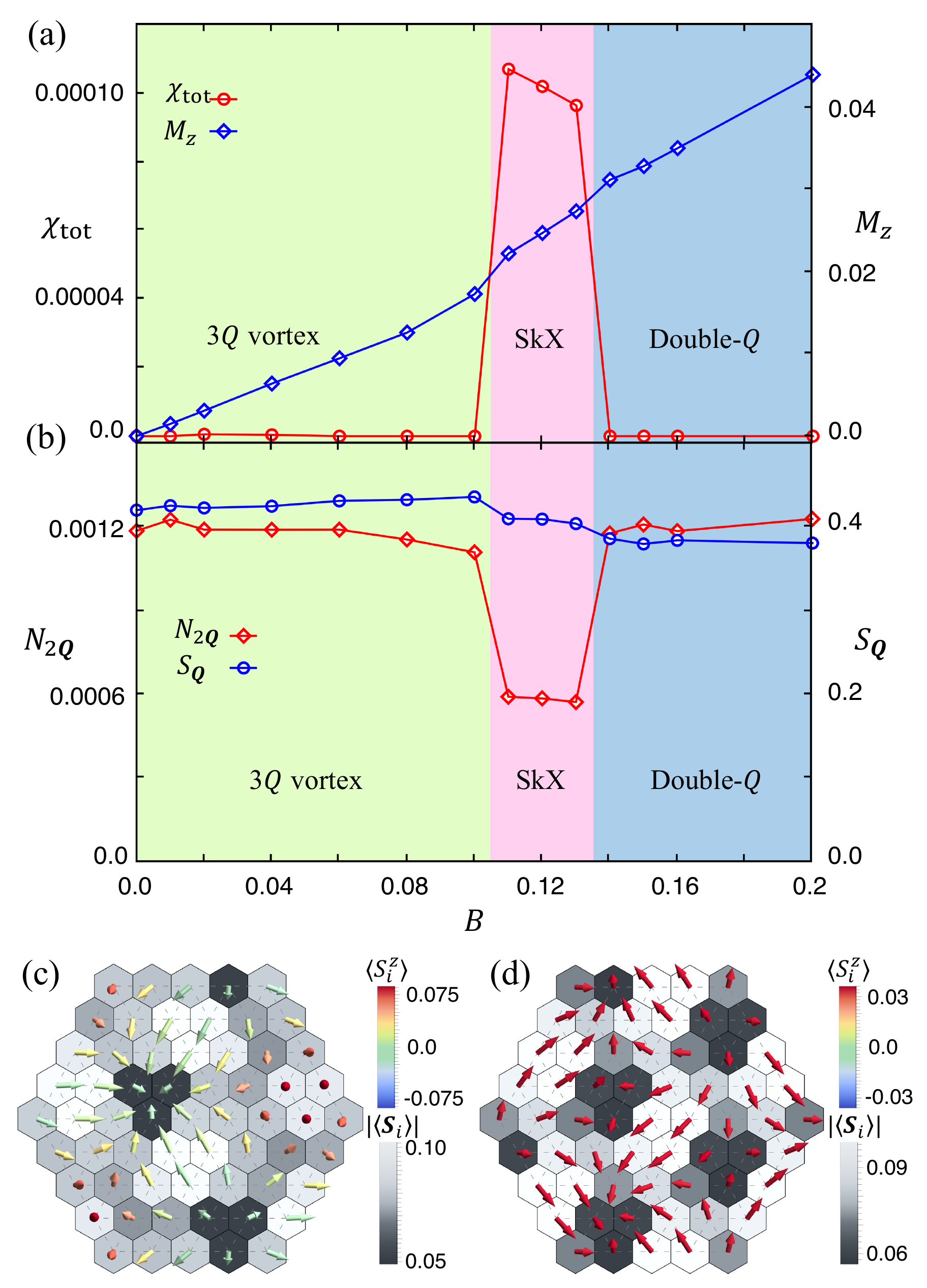}
  \caption{\(B\) dependence of (a) \(\chi_{\mathrm{tot}}\) and \(M_z\), (b) \(N_{2\bm{Q}}\) and \(S_{\bm{Q}}\). (c), (d) The spin configurations in (c) the SkX phase at $B=0.12$ and (d) the double-\(Q\) phase at $B=0.16$.
}
  \label{fig4}
\end{figure}

\begin{figure}[t]
  \includegraphics[width=\hsize]{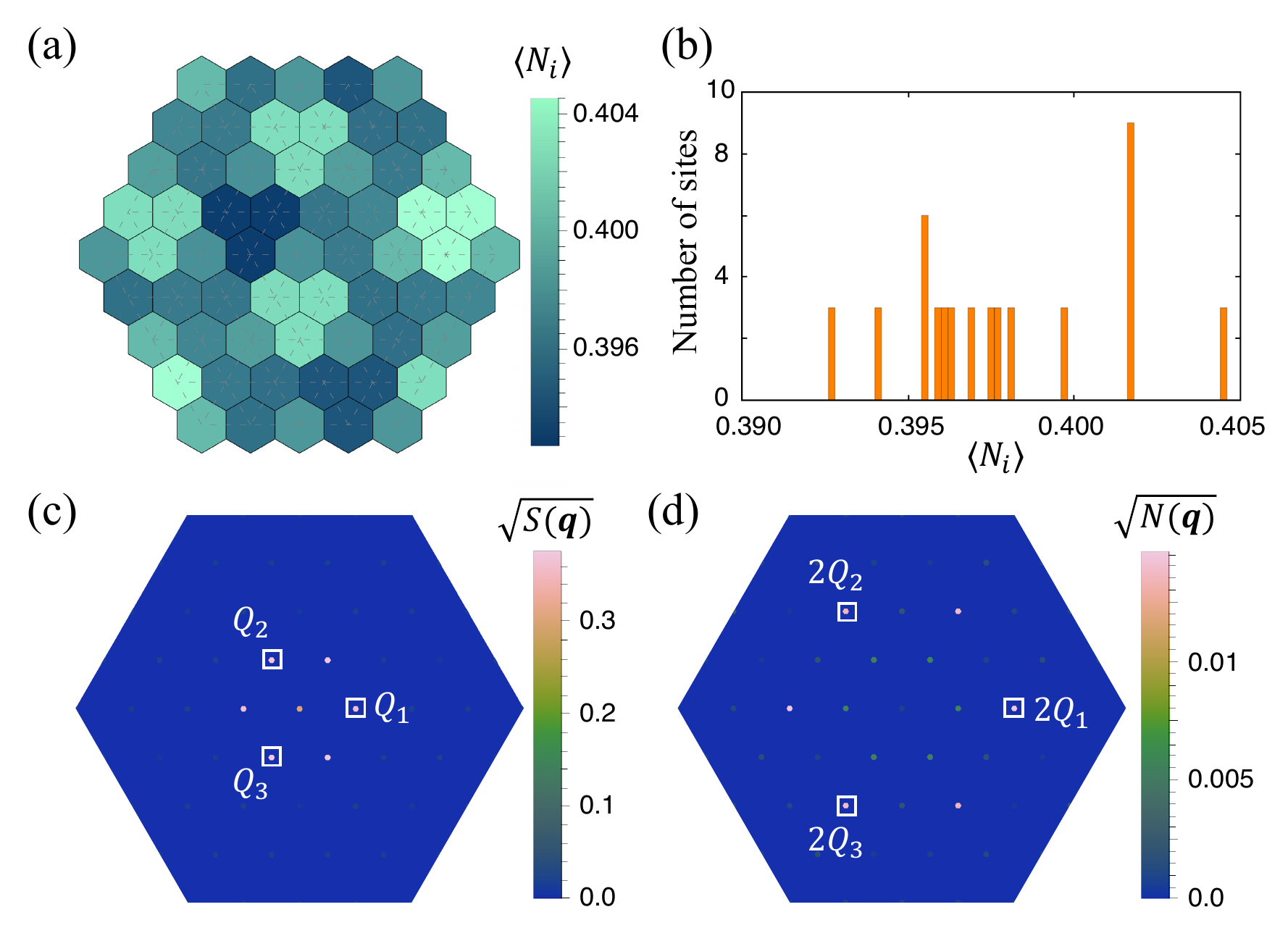}
  \caption{The data of the SkX phase at $B=0.12$. (a) The charge density distribution in real space. (b) The histogram of \(\langle N_i \rangle\) with a bin width of \(0.0002\). 
  (c),(d) The intensity plot of the square root of (c) \(S(\bm{q})\) and (d) \(N(\bm{q})\).}
  \label{fig5}
\end{figure}

Next, we consider the effect of the magnetic field $B$ on the 3$Q$ vortex phase. Figure~\ref{fig4}(a) shows the $B$ dependence of $\chi_{\mathrm{tot}}$ and $M_z$, while Fig.~\ref{fig4}(b) shows that of $N_{2\bm{Q}}$ and $S_{\bm{Q}}$ at $U=6.4$. When introducing $B$, $M_z$ becomes nonzero and linearly increases in the 3$Q$ vortex phase in Fig.~\ref{fig4}(a). While increasing $B$, $\chi_{\rm tot}$ jumps to nonzero values at $B=0.11$, which implies the appearance of the SkX phase. 
The obtained SkX spin configuration is shown in Fig.~\ref{fig4}(c). 
Compared with the SkX II in Fig.~\ref{fig2}(a), the spatial distributions of $|\langle \bm{S}_i\rangle |$ in Fig.~\ref{fig4}(c) and $\langle N_i \rangle $ in Fig.~\ref{fig5}(a) are more complicated; indeed, the histogram of $\langle N_i \rangle $ exhibits multiple values, as shown in Fig.~\ref{fig5}(b). 
Accordingly, additional intensities appear at $\bm{Q}_\mu$ and $\bm{Q}_\mu+\bm{Q}_{\mu'}$ in $N (\bm{q})$ in Fig.~\ref{fig5}(d). 
Owing to the multiple peak structures in $N (\bm{q})$, the intensity of $N_{2\bm{Q}}$ becomes small compared to that of the 3$Q$ vortex, as shown in Fig.~\ref{fig4}(b). 
In addition, this state exhibits a nonzero net magnetization owing to the magnetic field; see also the $\bm{q}=\bm{0}$ component in the spin structure factor in Fig.~\ref{fig5}(c). 
Thus, this SkX phase induced by the magnetic field is different from the SkX I and SkX II obtained at a zero field. 
Since a further increase of $B$ suppresses the \(z\)-spin modulation in the SkX, the spin state turns into the double-$Q$ state with the in-plane double-$Q$ modulated spin structure shown in Fig.~\ref{fig4}(d). 
In the double-$Q$ state, $N({\bm{q}})$ shows peaks at two out of three $2\bm{Q}_\mu$ like the 3$Q$ vortex, whose amplitude is comparable to each other [Fig.~\ref{fig4}(b)]. 

It is noteworthy to mention that the obtained SkX under the magnetic field is qualitatively different from the SkX found in the Kondo lattice model~\cite{2017:PRL:Ozawa} and other spin models. 
As closely looking into the spin configuration of the SkX phase in Fig.~\ref{fig4}(c), one finds that almost all the spins have a positive \(\langle S_i^z \rangle\) and only a few spins around the two skyrmion cores where $|\langle \bm{S}_i\rangle |$ becomes small have a slight negative \(\langle S_i^z \rangle\). These two skyrmion cores yield the winding number of \(+1\), i.e., the skyrmion number of $-0.5$, which reflects the slight negative \(\langle S_i^z \rangle\) at the core.
Intriguingly, in this phase, there is another vortex core, where the spin almost parallel to the \(+z\) direction in the middle-right side of the spin configuration.
This vortex accompanies the winding number of $-2$, i.e., the skyrmion number of $-1$. In the end, the total skyrmion number in the magnetic unit cell is given by 
$-2$.  
This is in contrast to the SkX in the Kondo lattice model, where $z$-spin component of the skyrmion core points along $S_i^z=-1$ and the spin configuration possesses the skyrmion number of $-1$~\cite{2017:PRL:Ozawa}.

To summarize, we have studied the ground-state spin configurations in the Hubbard model on the centrosymmetric triangular lattice with an emphasis on the itinerant nature of electrons. 
Based on the mean-field analysis, we have discovered several multiple-\(Q\) phases including the SkX and vortex phases in the presence of the strong spin-charge entanglement. Our result indicates that itinerant magnets provide a variety of topological spin crystals even without the classical localized spins, which provides a further possibility to search for exotic topological magnetism.

We thank R. Yambe and M. Yatsushiro for fruitful discussions. This research was supported by JSPS KAKENHI Grants Numbers JP21H01037, JP22H04468, JP22H00101, JP22H01183, and by JST PRESTO (JPMJPR20L8). 

\bibliography{bibtex} 

\end{document}